\documentclass[%
 aip,
 rsi,
amsmath,amssymb,
reprint,
floatfix
]{revtex4-1}

\usepackage{graphicx}
\usepackage{float}
\usepackage{dcolumn}
\usepackage{bm}

\usepackage[utf8]{inputenc}
\usepackage{mathptmx}
\usepackage{braket}
\usepackage[english]{babel}
\usepackage{xcolor}
\colorlet{RED}{red}
\colorlet{BLUE}{blue}
\usepackage{dcolumn}
\usepackage{bm}
\usepackage[version=4]{mhchem} 
\usepackage{acronym}
\usepackage{adjustbox}
\usepackage{tikz}
\usepackage{microtype} 
\usetikzlibrary{calc,shapes.geometric,decorations.pathmorphing,patterns}

\definecolor{background-color}{gray}{0.98}
\usepackage[margin=2.3cm,bmargin=1cm,footnotesep=1cm]{geometry}

\newcommand{\QuArC}{%
    \affiliation{%
        Microsoft Quantum,
        Redmond WA 98052, USA.}}
\newcommand{\PNNL}{%
    \affiliation{%
        Physical Sciences Division, Battelle, 
        Pacific Northwest National Laboratory, K8-91, P.O. Box 999, Richland WA 99352, USA}}
\newcommand{\PNNLHPC}{%
    \affiliation{%
        Advanced Computing, Mathematics, and Data Division, Battelle,                                  
        Pacific Northwest National Laboratory, K8-91, P.O. Box 999, Richland WA 99352, USA}}

\begin{document}

\title{Towards quantum computing for high-energy excited states in molecular systems: quantum phase estimations of core-level states}

\author{Nicholas P. Bauman} \PNNL
\author{Hongbin Liu} 
\email{hongbin.liu@microsoft.com} \QuArC
\author{Eric J. Bylaska} \PNNL
\author{Sriram Krishnamoorthy} \PNNLHPC
\author{Guang Hao Low} \QuArC
\author{Christopher E.  Granade} \QuArC
\author{Nathan Wiebe} \PNNL
\author{Nathan A. Baker} \PNNL
\author{Bo Peng} \PNNL
\author{Martin Roetteler} \QuArC
\author{Matthias Troyer} \QuArC
\author{Karol Kowalski} 
\email{karol.kowalski@pnnl.gov}
\PNNL

\date{April 2020}

\date{\today}

\begin{abstract}
This paper explores the utility of the quantum phase estimation (QPE) in calculating high-energy excited states characterized by promotions of electrons occupying inner energy shells.
These states have been intensively studied over the last few decades especially in supporting the experimental effort at light sources. 
Results obtained with the QPE are compared with various  high-accuracy many-body techniques developed to describe core-level states. 
The feasibility of the quantum phase estimator in identifying classes of challenging shake-up states characterized by the presence of higher-order excitation effects is also discussed. 
\end{abstract}

\maketitle

\section{Introduction}
The development of reliable theoretical modeling tools for describing excited states of complex molecules and molecular assemblies are central to advancing several science domains that span chemistry, physics, materials science, and biology.
These methodologies play key roles in understanding processes associated with the control of energy  capture and transfer through photochemical processes in a broad class of light harvesting systems,\cite{mozer2006conjugated,li2008conjugated,coakley2004conjugated} photocatalytic hydrogen production from water,\cite{sprick2015tunable} facilitating proton coupled transfer in redox reactions enabling water oxidation,\cite{maneiro2003kinetics,keough2011proton,yamaguchi2014regulating} carrier dynamics in nanoparticles and materials,\cite{kongkanand2008quantum,tachikawa2007mechanistic,hartland2011optical} photoactivation processes in proteins,\cite{crosson2001structure,harper2003structural,beja2001proteorhodopsin} bioluminescence of living organisms,\cite{rees1998origins,tsien1998green} and ultra-fast protective mechanisms in DNA.\cite{peon2001dna,pecourt2001dna}
Additionally, the rapid development of highly tunable advanced light sources has enabled a wide spectrum of various X-ray spectroscopies.\cite{chergui2016time,young2018roadmap,stohr2013nexafs,de2008core,milne2014recent,chergui2017photoinduced}  
Core-level (CL) spectroscopies like X-ray absorption (XAS), X-ray emission (XES), resonant inelastic X-ray scattering (RIXS), X-ray magnetic circular dichroism (XMCD) and X-ray photoelectron (XPS),  have significantly advanced our understanding of the structure and properties of matter.
Generally  speaking, XAS probes the electronic and geometric structure of matter at the atomic level by recording excitations from a core-level
state of an atom to either bound valence or continuum states, resulting in energetically distinct absorption edges.

Over the last few decades, these experimental advances have triggered a significant effort towards enabling accurate theoretical methods for X-ray spectroscopy ranging from time-dependent density functional theory (TD-DFT) to Green's function theory and  wave-function-based configuration interaction (CI)  and coupled cluster (CC) methods.
\cite{fadley1969multiplet,svensson1976scf,aagren1975scf,cederbaum1980many1,cederbaum1980many2,barth1985theoretical,zabinsky1995multiple,ankudinov1998real,rehr2000theoretical,nooijen1995description,ohtsuka2006inner,kong2012roles,leng2016gw,norman2018simulating,oosterbaan2020generalized,bokarev2020theoretical}
Although CC methods are more expensive, they are systematically improvable by including higher-rank collective excitation effects in the wave function expansion.
Owing to this property, several CC methodologies for core-level states have been intensively developed including single- and multi-reference CC formulations.\cite{coriani2012coupled, sonia2015cvs,myhre2016near,peng2015energy,sen2013study,brabec_core,nascimento2017simulation,sadybekov2017coupled,park2019equation,lee2019excited,tenorio2019molecular,carbone2019analysis,matthews2020eom,vidal2020dyson,rehr2020equation}

Linear response \cite{monkhorst77_421,jorgensen90_3333} or closely related  equation-of-motion coupled cluster \cite{bartlett89_57,bartlett93_414,stanton93_7029} methods
(LR-CC/EOMCC) have evolved into the most accurate theoretical techniques to describe excited-state processes; however, their application is significantly limited by the flaws of numerical procedures to directly approach states in the core-level regime.
Several methods have been proposed to address these issues, including Lanczos algorithms,\cite{coriani2012asymmetric} reduced-space algorithms to solve the complex coupled cluster linear response equations of damped response theory,\cite{kauczor} and core-valence separation methods.\cite{sonia2015cvs,vidal2020dyson}
However,  these problems scale with the  increasing rank  of EOMCC approximations \cite{hirata2004higher} needed to provide the desired level of accuracy especially in the context of states dominated by higher-than-single excitations, which is a typical situation in describing broad class of shake-up/satellite states. 

This paper addresses the above-mentioned  problems associated with the accuracy and the  identifiability  of  classes of solutions of Schr\"odinger equation corresponding to high-energy excited states with predefined configurational structures.
In particular, we should how these problems can be addressed using quantum computing and quantum phase estimation algorithms (QPE) \cite{Kitaev:97,NC:2000}.
We focus our analysis on the example of core-level states of the water molecule, where the core-level states of various spin multiplicities, excitation levels, and spatial symmetries are investigated.
Particular attention is paid to model doubly (and higher) excited core-level state, which usually pose significant problems for low-rank EOMCC approximations. 
\section{Many-body formulations for core-level states}
We compare QPE results obtained with small active spaces that allow the use of quantum simulators to results obtained with with two other methods.
Comparison is made with EOMCC methods\footnote{EOMCCSD \cite{bartlett93_414} and  EOMCCSDT \cite{kkppeom} models including single and double, and single, double, and triple excitations respectively}.
Additionally, the QPE results are compared with state-specific multi-reference coupled cluster (MRCC) formulations based on Brillouin-Wigner (BW-MRCC) \cite{pittnermasik,bwpittner1} and Mukherjee's (Mk-MRCC) approaches.\cite{mahapatra1,mahapatra2,evangelista1,evangelista2,daskallay}
The determinant-based full configuration interaction (FCI) \cite{knowles89_75} has been applied to the same active space as a test of QPE simulation.

The general form of the ansatzes for $K$-th electronic state $|\Psi_K\rangle$ employed by  EOMCC and MRCC formulations take the form
\begin{eqnarray}
  |\Psi_K^{\rm EOMCC}\rangle &=& R_K e^{T} |\Phi\rangle \;, 
  \label{eomcc} \\
  |\Psi_K^{\rm MRCC}\rangle &=&  \sum_{\mu=1}^{M}
  c_{\mu}(K) e^{T_{\mu}(K)}  |\Phi_{\mu}\rangle  \;,
  \label{mrcc}
\end{eqnarray}
where (1) $R_K$ is the state-specific excitation operator for $K$-th state, $T$ is a cluster operator obtained in standard ground-state CC calculations, and $|\Phi\rangle$ stands for the so-called reference function (usually chosen as a Hartree-Fock (HF) determinant), (2) $c_{\mu}(K)$ are components of the eigenvector of the effective Hamiltonian diagonalized in the active space, $T_{\mu}(K)$ are reference-specific cluster operators describing $K$-th state, and Slater determinants $|\Phi_{\mu}\rangle$ span the so-called model space (${\cal M}_0$; ${\rm dim} ({\cal M}_0)=M$). 
To determine amplitudes defining $T$/$R_K$ and $T_{\mu}(K)$ operators one solves standard CC/EOMCC equations and  Brillouin-Wigner/Mukherjee sufficiency conditions. 

In order to  numerically   identify  core-level states with the EOMCC methodologies, one needs an accurate initial guess for the EOMCC diagonalization  and efficient iterative diagonalization procedures. 
In some cases, as discussed in Ref. \onlinecite{brabec_core}, meeting both requirements may pose a significant challenge. 
Calculations with MRCC methods require a judicious choice of active space (or equivalently active orbitals).
In the present study, we compare QPE results with the MRCCSD (MRCC with singles and doubles)  results obtained with ${\cal M}_0$ containing relevant active core orbitals and active valence unoccupied $V_u^{CL}$ (${\cal M}_0^{\rm CL}$) for core-level states and active-space valence occupied/unoccupied  orbitals $V_o^G$/$V_u^G$ (${\cal M}_0^{\rm G}$) for the ground-state calculations.
The MRCCSD CL excitation energies are formed as a difference of two independent state-specific MRCCSD calculations
\begin{equation}
  \omega^{\rm CL}=E_{\rm MRCC}^{\rm CL}({\cal M}_{0}^{\rm CL}) - E_{\rm MRCC}^{\rm G}({\cal M}_{0}^{\rm G})\;.
  \label{eene}
\end{equation}
As discussed in Ref. \onlinecite{brabec_core}, EOMCC and MRCC formalisms can be used to provide CL excitation energies in a good agreement with the experimental values. 
%
\section{Quantum Algorithms: Quantum Phase Estimation}
Quantum phase estimation (QPE) allows one to estimate the eigenvalue $\lambda$ corresponding to an eigenvector $\ket{\psi_\lambda}$ of a given operator $U$; i.e., $U \ket{\psi_\lambda} = \lambda \ket{\psi_\lambda}$. 
The distribution of energies for the ground and excited states from the QPE algorithm is determined by the Hamiltonian and a trial wavefunction composed of a superposition of Slater determinants, wherein the probability of obtaining an energy estimate for a particular state is proportional to the amount of overlap of the trial wave with that corresponding eigenstate. %
Through repeated simulations, one accumulates samples from this distribution of eigenstate energies.
The error in each energy estimate is inversely proportional to the number of applications of the time evolution operator $U=e^{-iH\Delta}$ in the QPE algorithm -- specified either through the number of ancillary qubits used in QPE, or the targeted bits-of-precision in the robust phase estimation (RPE) variant that uses only one ancillary qubit.

The QPE method is in contrast to variational quantum eigensolver (VQE) approaches, which only provide energy estimates for a single targeted state. 
Moreover, the error in VQE estimate is uncontrolled -- one only knows that it upper-bounds that energy of the ground state, but the gap between the target energy and the obtained estimate is unknown. 

Thus QPE approaches open up opportunities to find and chronicle exotic and novel states that are unobtainable with conventional computing and current approximate formulations.
We recommend readers Refs. \onlinecite{Kitaev:97} and \onlinecite{NC:2000} for additional details of QPE, and Ref. \citenum{KLY:2015} for RPE, which is used in our later numerical experiments. 
Both approaches have been implemented within the Microsoft Quantum Development Kit (QDK),\cite{Svore18_10} and the following code listing in the Q\# programming language effectively demonstrates how the algorithm is implemented.
Here, a given unitary operation \texttt{U} acts on a quantum register \texttt{psi}. The estimate of the eigenvalue is stored in a control register \texttt{ctrlReg}.
{\small
\begin{verbatim}
operation QPE (U : DiscreteOracle, psi : Qubit[], 
  ctrlReg : BigEndian) : Unit is Adj + Ctl {
  
  let nQubits = Length(ctrlReg!);
  AssertAllZeroWithinTolerance(ctrlReg!, 1E-10);
  ApplyToEachCA(H, ctrlReg!);

  for (index in 0 .. nQubits - 1) {
    let control = (ctrlReg!)[index];
    let power = 2 ^ ((nQubits - index) - 1);
    Controlled U!([control], (power, psi));
  }
  
  Adjoint QFT(ctrlReg);
}
\end{verbatim}
}
We omit the implementation of RPE, as it is somewhat more involved, but include for completeness the code listing that invokes it
{\small
\begin{verbatim}
operation RobustPhaseEstimation (
  bitsPrecision : Int, oracle : DiscreteOracle, 
  targetState : Qubit[]) : Double
{ ... }
\end{verbatim}
}

Given a target error of $\epsilon$ (expressed in units of Hartree), the number of times $U=e^{-iH\Delta}$ is applied in the phase estimation algorithm scales with $\mathcal{O}(1/(\epsilon \Delta))$ for each energy estimate.
In this manuscript, the quantum circuit that approximates $U$ is obtained by a first-order Trotter algorithm.
As detailed in Ref. \citenum{low2019q}, the electronic structure Hamiltonian is represented in terms of Fermion operators in the second-quantized representation with $N$ orbitals is encoded into a sum of Pauli operators on $2N$ qubits by the Jordan-Wigner representation. 
In this representation, $H=\sum_{j=0}^{M-1}\alpha_j P_j$, where $\alpha_j$ are real coefficients, $P_j$ are Pauli operators, and the number of terms $M$ scales with $\mathcal{O}(N^4)$.
Time-evolution by $H$ is then approximated by applying time-evolution of its component terms in succession like $e^{-iH\Delta}=\prod_{j=0}^{M-1}e^{-i\alpha_j P_j\Delta}+\mathcal{O}(\Delta^3\sum_{j,k,\ell}\alpha_j\alpha_k\alpha_\ell\|[P_j,[P_k.P_\ell]]\|)$.
This equation provides some guidance on how the step size $\Delta$ should be chosen.
As the number of exponentials required is $M=\mathcal{O}(N^4)$ and each Pauli operator is of length $\le 2N$, $U$ may be realized using at most $\mathcal{O}(N^5)$ quantum gates. 
Using the optimizations described in Ref. \citenum{wecker2015progress}, the number of quantum gates may be reduced to $\mathcal{O}(N^4)$ by instead paying an amortized cost of $\mathcal{O}(1)$ for these Pauli operators. 
Our error bounds suggest that, because there are $N^{10}$ commutators in the error expansion, if we take the Hamiltonian coefficients to be upper bounded by a constant, then $\Delta \in \mathcal{O}(\epsilon/N^{10/3})$ suffices for the Phase estimation.
Our bounds on the cost of performing phase estimation then yield $\mathcal{O}(N^{22/3}/\epsilon)$ scaling using this approach are therefore in $\mathcal{O}(N^4/(\Delta \epsilon) \subseteq \mathcal{O}(N^{22/3}/\epsilon)$.
As the empirically obtained error is often much smaller than predicted by this upper bounds, we choose $\Delta=0.1$ in this manuscript which suffices to distinguish the different electronic transitions that we consider. 
\section{Computational Details}
Qe chose the water molecule as described by the cc-pVDZ basis set\cite{dunning89_1007} as a benchmark system for classical and quantum simulations.
This system has been  a subject of the EOMCC and MRCC studies of core-level states dominated by the single excitations of the electron from the $1s$ orbital of the oxygen atom.
The geometry of the system is given by $R_{\rm O-H}=0.9772$\AA\; and $\theta_{\rm H-O-H} = 104.52\deg$.
The same geometry has been used in the EOMCCSD, EOMCCSDT, Mk-MRCCSD calculations reported in Ref. \onlinecite{brabec_core}.
The QPE simulation and FCI calculation are conducted using a truncated 9 orbitals active space that corresponds to the 9 lowest-lying orbitals ($1a_1$, $2a_1$, $1b_1$, $3a_1$, $1b_2$, $4a_1$, $2b_1$, $2b_2$, $3b_2$ orbitals) and considering all electrons as being correlated.
\section{Results}
Singly excited core-level states corresponding to the $1a_1\rightarrow 4a_1$ and $1a_1\rightarrow 2b_1$ transitions have been observed experimentally\cite{schirmer1993k} and benchmarked with single- and multi-reference \cite{brabec_core, wight1974k} as summarized in Table \ref{table_h2o_cl}.
These states are easily described by the QPE simulation employing active space defined by 9 lowest-lying Hartree-Fock orbitals, when either excitation is used as an initial guess.
At first glance, the quantum simulations appears to overestimate the excitation energies for both transition by $\sim10$ eV, when compared with the best CC results.
This discrepancy is a consequence of missing dynamical correlation outside of the active space, a limitation due to computational limitations of the simulations.
Instead, it is more appropriate to compare with the FCI energies obtained with the same truncated (active) orbital space.
When this is done, the QPE results are in a very good agreement with the corresponding FCI excitation energies.
It is rational to assume that much of the same correlations are missing for both states.
This is reflected in the relative energies between these two states, which is $\sim2$ eV, and essentially the same difference observed with experiment and CC benchmarks.

Our earlier study of excited states with the QPE\cite{Bauman19_234114} demonstrated that the stochastic nature of the QPE can facilitate the discovery/identification of excited states or excited-state processes in situations when the knowledge about the true configurational structure of a sought after excited state is limited or postulated. 
We explore this idea by investigating double excited core-level states of H$_2$O, which can plague or elude conventional many-body methods.
As an example, we chose the state/states of the $^1B_2$ symmetry that are dominated by $|\Psi_{\rm ini}\rangle = \frac{1}{\sqrt{2}} (|\Phi_{1\bar{5}}^{6\bar{6}}\rangle + |\Phi_{5\bar{1}}^{6\bar{6}}\rangle$ (see Fig.~\ref{fig1}) singlet combination of doubly excited Slater determinants. 
This combination was also employed as an initial guess for QPE simulations. 
Of the 200 simulations, we focused on the most prominent states, which are the two lowest-energy doubly excited core-level states in this series, seen in Fig.~\ref{figshakeup}.
These two states, with total energies of -55.3088$\pm$0.0039 and -55.2475$\pm$0.0038 Hartree, are confirmed by independent FCI calculations to be the lowest-energy singlet states with leading $|\Phi_{1\bar{5}}^{6\bar{6}}\rangle$ and $|\Phi_{5\bar{1}}^{6\bar{6}}\rangle$ excitations.
The total energies obtained with FCI are -55.3088 and -55.2472 Hartree, respectively, demonstrating that QPE simulations can effectively capture complex excited states when using a postulated initial guess. 

QPE simulations were also performed for triplet core-level state of $A_1$ and $B_1$ symmetries.
The results of QPE and FCI simulations are shown in Table~\ref{table_triplets}.
In analogy to the singlet case, excitations energies of core-level $^3A_1$ and $^3B_1$ states obtained with QPE in the active space are in excellent agreement with the FCI excitation energies obtained for the same active space.
The comparison with the HF results in Table~\ref{table_triplets} emphasizes the role played by the correlation effects in the proper description of these states. 

\renewcommand{\tabcolsep}{0.2cm}
\begin{center}
  \begin{table*}
    \centering
    \caption{Excitation energies (in electron volts (eV)) of low-lying singly excited core-level states in the H$_2$O molecule described by the cc-pVDZ basis set.\cite{dunning89_1007}
    Experimental geometry 
    ($\theta_{\rm H-O-H} = 104.52\deg \;\; 
    R_{\rm O-H}=0.9772$ \AA)
    was used in all calculations.\\}
    \begin{tabular}{lcccccccc} \hline \hline  \\
    Transition & HF\protect\footnotemark[1] & EOMCCSD\protect\footnotemark[2] & EOMCCSDT\protect\footnotemark[2] & BW-MRCCSD\protect\footnotemark[2] & Mk-MRCCSD\protect\footnotemark[2] & QDK (10,9)\protect\footnotemark[3] & FCI(10,9) & Expt.\protect\footnotemark[4] \\[0.1cm]
    \hline \\
    $1a_1\rightarrow 4a_1$ & 564.23 & 538.40 & 537.32 & 537.56 & 537.62 & 547.15 & 547.19 &
    534.0 \\[0.3cm]
    $1a_1\rightarrow 2b_1$ & 566.23 & 540.21 & 539.26 & 539.49 & 539.55 & 549.14 & 549.15 & 
    535.9 \\[0.2cm]
    \hline \hline 
    \end{tabular}
    \label{table_h2o_cl}
    \footnotetext[1]{
    Differences of HF orbital energies corresponding to leading excitations.
    }
    \footnotetext[2]{
    Results taken from Ref. \onlinecite{brabec_core}.
    }
    \footnotetext[3]{
    QDK calculations were performed for 10 electrons in 9 lowest-lying molecular HF orbitals. 
    Excitation energies are reported as differences between averaged energy values for ground 
    and core-level states, i.e., $-76.0591\pm 0.0041$, $-55.9517\pm0.0040$, 
    and $-55.8785\pm 0.0042$ Hartree.
    }
    \footnotetext[4]{Experimental values of excitation energies taken from Ref. \onlinecite{wight1974k}.
    }
  \end{table*}
\end{center}
%
\begin{widetext}
  \begin{center}
    \begin{figure}[H]
      \includegraphics[trim={0 0.5cm 0 0},clip,width=0.9\textwidth]{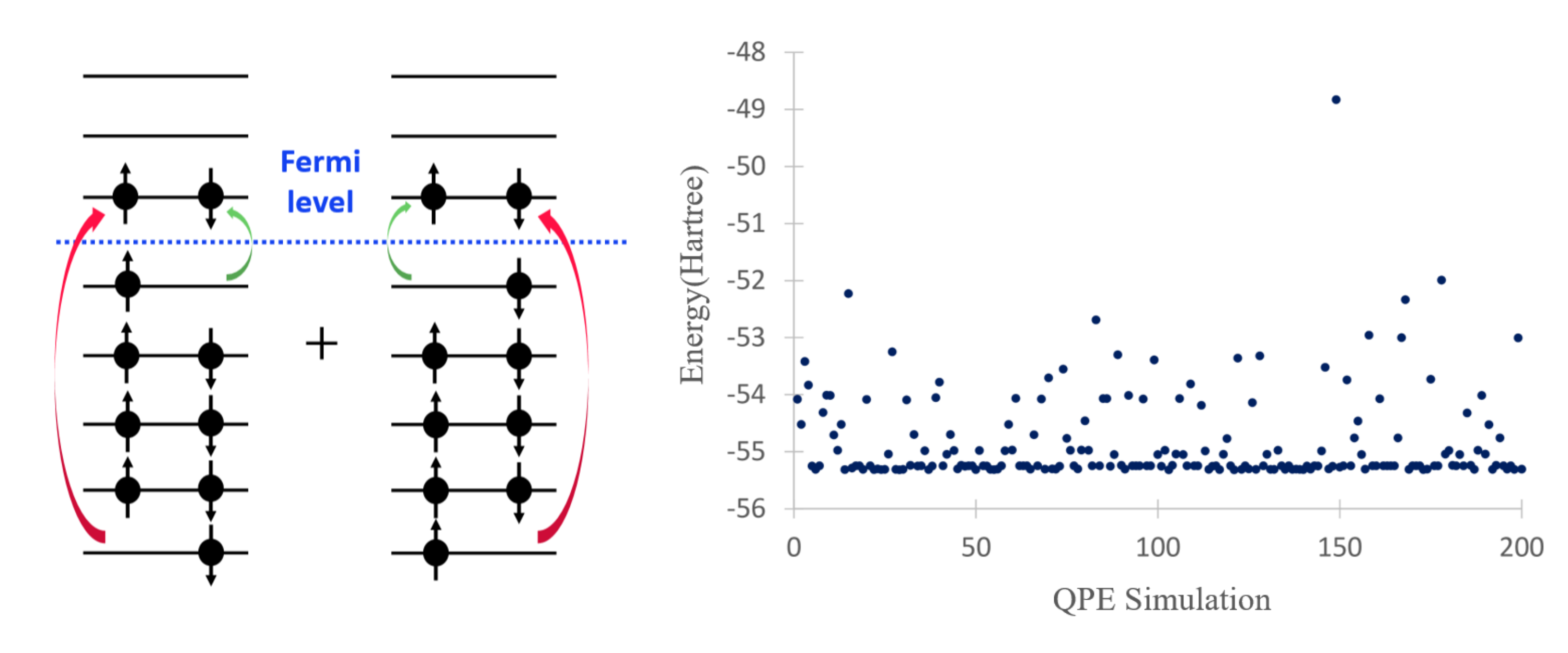}
      \caption{Schematic representation of QPE simulations of doubly excited singlet core-level states of the  H$_2$O system in cc-pVDZ basis set \cite{dunning89_1007} using initial guess representing singlet combination of Slater determinants corresponding to simultaneous excitation of electrons from core and valence levels to virtual valence level 
      ($|\Psi_{\rm ini}\rangle = \frac{1}{\sqrt{2}} 
      (|\Phi_{1\bar{5}}^{6\bar{6}}\rangle + 
      |\Phi_{5\bar{1}}^{6\bar{6}}\rangle$).}
      \label{fig1}
    \end{figure}
  \end{center}
\end{widetext}
\begin{center}
  \begin{figure}[H]
    \includegraphics[trim={2.6cm 3.6cm 2cm 3.1cm},clip,width=0.48\textwidth]{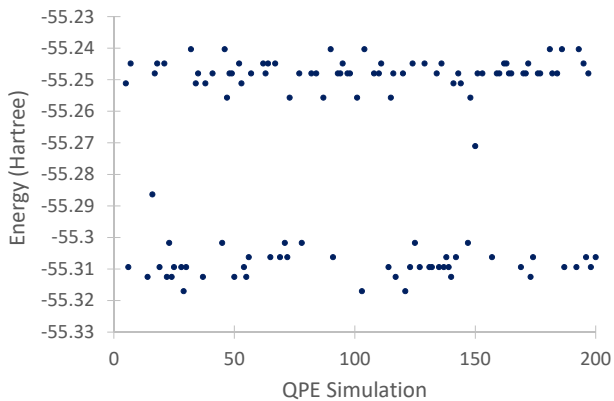}
    \caption{QPE simulations of the two lowest-energy doubly excited core-level states of the  H$_2$O system in cc-pVDZ basis set using initial guess representing singlet combination of Slater determinants corresponding to simultaneous excitation of electrons from core and valence levels to virtual valence level 
    ($|\Psi_{\rm ini}\rangle = \frac{1}{\sqrt{2}} 
    (|\Phi_{1\bar{5}}^{6\bar{6}}\rangle + 
    |\Phi_{5\bar{1}}^{6\bar{6}}\rangle$).}
    \label{figshakeup}
  \end{figure}
\end{center}
\renewcommand{\tabcolsep}{0.2cm}
\begin{center}
  \begin{table*}
    \centering
    \caption{Excitation energies (in electron volts (eV)) of low-lying singly excited triplet core-level states in the H$_2$O molecule described by the cc-pVDZ basis set.\cite{dunning89_1007}
    Experimental geometry 
    ($\theta_{\rm H-O-H} = 104.52\deg \;\; 
    R_{\rm O-H}=0.9772$ \AA)
    was used in all calculations.\\}
    \begin{tabular}{lccc} \hline \hline  \\
    Transition & HF\protect\footnotemark[1] & QDK (10,9)\protect\footnotemark[1] & FCI(10,9) \\[0.1cm]
    \hline \\
    $1a_1\rightarrow 4a_1$ & 564.23 & 546.72 & 546.81 \\[0.3cm]
    $1a_1\rightarrow 2b_1$ & 566.23 & 548.95 & 548.96 \\[0.2cm]
    \hline \hline 
    \end{tabular}
    \footnotetext[1]{
    Differences of HF orbital energies corresponding to leading excitations.
    }
    \footnotetext[2]{
    QDK calculations were performed for 10 electrons in 9 lowest-lying molecular HF orbitals. 
    Excitation energies are reported as differences between averaged energy values for ground 
    and core-level states, i.e., $-76.0591\pm 0.0041$, $-55.9674\pm0.0042$, and
    $-55.8855\pm 0.0039$ Hartree.
    }
    \label{table_triplets}
  \end{table*}
\end{center}
%
%
\section{Conclusions}
We have demonstrated that QPE algorithm can be extended to describe high-energy core-level states corresponding to  various spin and spatial symmetries.
For all core-level states of the water molecule considered here, the QPE results successfully reproduced the classical FCI ones obtained in the same active space.
We also demonstrated that QPE can ``echo'' states that are statistically relevant to  (or have non-negligible overlap with)  the specific hypothesis state defined by the initial state.
This was demonstrated on the example of doubly excited states of H$_2$O of the $^1B_2$ symmetry.
It is important to emphasize that these states usually require high-rank excitations that need to be included in the CC/EOMCC cluster and excitation operators, which in many cases makes the task of identifying these states very challenging.
Summarizing, the examples considered in this paper provide a good illustration of the universal character  of the QPE algorithm for  locating electronic excited states across various energy scales. 
\section{Acknowledgement}
This  work  was  supported  by  the ``Embedding Quantum Computing into Many-body Frameworks for Strongly Correlated  Molecular and Materials Systems'' project, which is funded by the U.S. Department of Energy(DOE), Office of Science, Office of Basic Energy Sciences, the Division of Chemical Sciences, Geosciences, and Biosciences.
This research was also funded by the ``Quantum Algorithms, Software, and Architectures (QUASAR)'' Agile Investment at Pacific Northwest National Laboratory (PNNL).
It was conducted under the Laboratory Directed Research and Development Program at PNNL.
QPE simulations and related EOMCC calculations have been performed using the Molecular Science Computing Facility (MSCF) in the Environmental Molecular Sciences Laboratory (EMSL) at the Pacific Northwest National Laboratory (PNNL).
FCI calculations were performed on the cloud computing infrastructures of Microsoft Azure.
EMSL is funded by the Office of Biological and Environmental Research in the U.S. Department of Energy.
PNNL is operated for the U.S. Department of Energy by the Battelle Memorial Institute under Contract DE-AC06-76RLO-1830.


%

\end{document}